\documentclass{elsart}
\usepackage{natbib}
\usepackage{epsfig}

\begin{document}
\runauthor{Tasitsiomi, Gaskins and Olinto}

\begin{frontmatter}

\title{Neutralino annihilation $\gamma$-rays from clumps and the LMC}
\author[astro,cfcp]{Argyro Tasitsiomi\thanksref{email}},
\author[physics,cfcp]{Jennifer Gaskins}, 
\author[astro,cfcp]{Angela V. Olinto }
\address[astro]{Department of Astronomy and Astrophysics, and} 
\address[physics]{Department of Physics, and}
\address[cfcp] {Center for Cosmological Physics, 
The University of Chicago, Chicago IL 60637}
\thanks[email]{iro@oddjob.uchicago.edu}


\begin{abstract}
We discuss the detectability of dark matter clumps in the Milky Way halo
due to neutralino annihilation. We then focus on a known ``clump'', the
Large Magellanic Cloud (LMC).
\end{abstract}


\end{frontmatter}

\section{Dark matter clumps}
\label{sec:clumps}
\begin{figure}
\epsfig{file=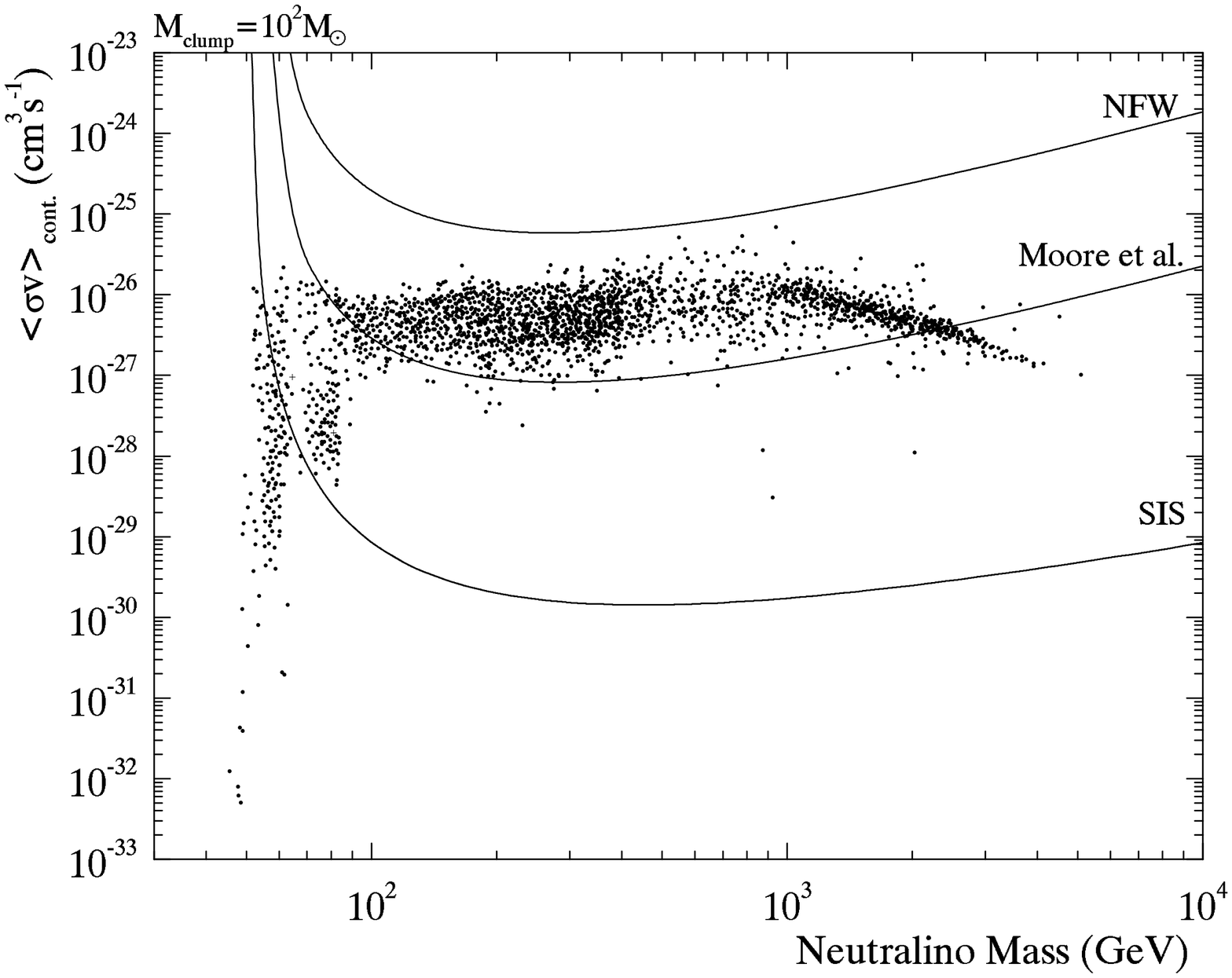, width=7.3cm, height=6.4cm}
\epsfig{file=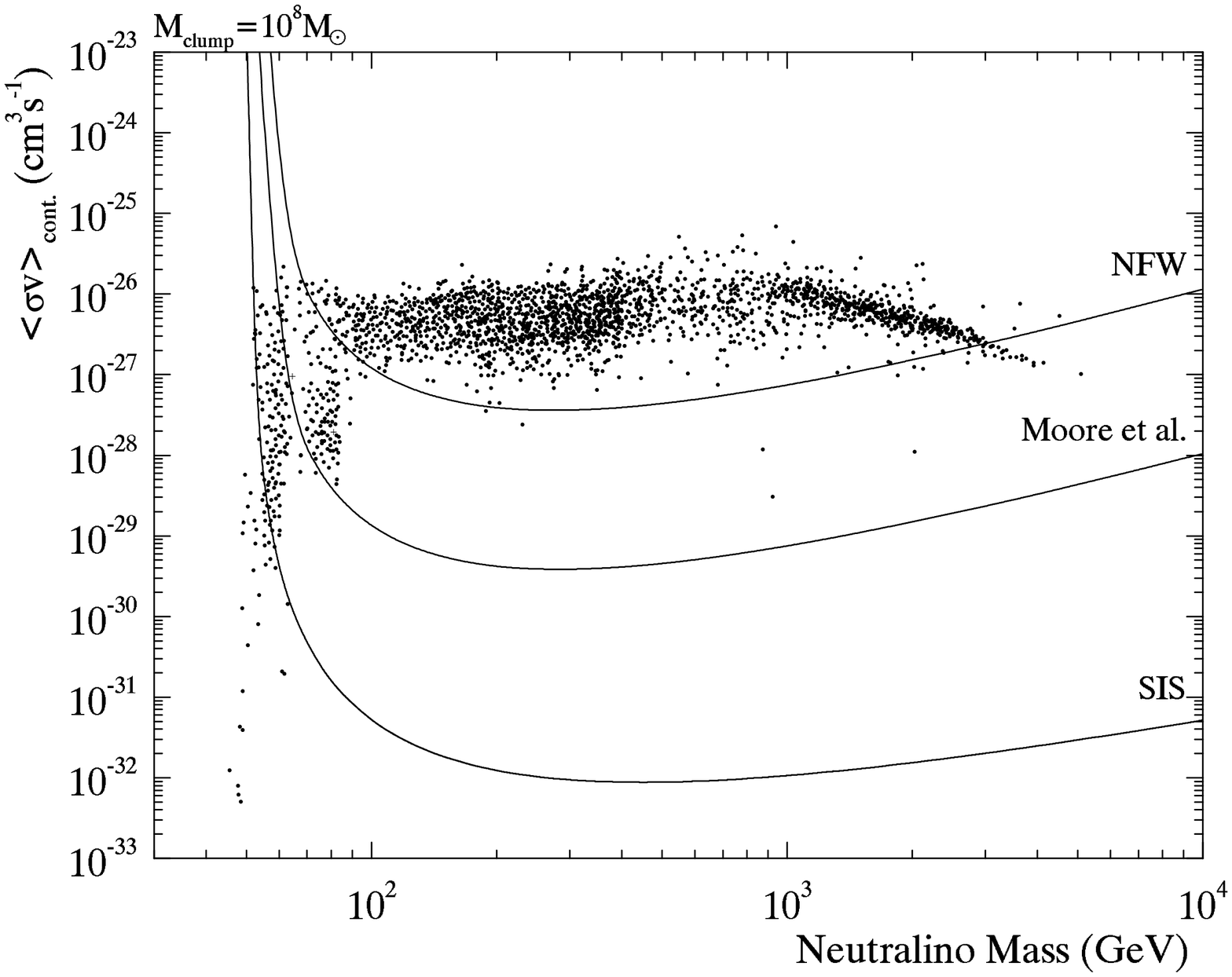, width=7.3cm, height=6.4cm}
\caption{\label{conti-susy}
The minimum detectable $\langle \sigma v \rangle_{cont.}$  versus
$m_{\chi}$ for the SIS, the Moore et al., and the NFW profile.
The clump masses used are $10^{2}
M_{\odot}$ (left panel) and $10^{8} M_{\odot}$ (right panel).
The dots represent possible SUSY models. The lines represent  the
5-$\sigma$  detection limits. 
Only  SUSY models that lie above the corresponding curve will
yield a detectable signal. }
\end{figure}

High resolution N-body simulations have revealed the survival of
considerable substructure  
within galactic halos. 
Assuming these substructure clumps are composed of annihilating neutralinos, 
the flux $F$ 
of a clump at distance $d$  with a density
distribution $\rho(r)$ is 
\begin{equation}
F=\frac{1}{2} \frac{1}{4 \pi d^{2}} \frac{N_{\gamma} \langle \sigma v \rangle}{m_{\chi}^{2}}
\int_{0}^{R} \rho^{2}(r) d^{3}r,
\end{equation}
where $\langle \sigma v \rangle$ is  the thermally averaged annihilation cross section,
$m_{\chi}$ is the neutralino mass, and $N_{\gamma}$ is the number of photons per annihilation 
with energy above an assumed energy threshold. 
For the density profile of the clumps we use the
\citet{navarro_etal96} (NFW) and the 
\citet{moore_etal98} profiles which were  
found to describe adequately the dark matter halos  in simulations. As
an upper limit to the
degree of central concentration of the actual clump profile we also consider the singular 
isothermal sphere (SIS).  
The three profiles behave at the center as $r^{-1}$, $r^{-1.5}$,  and $r^{-2}$, 
respectively. The results for the profiles and for the minimum and maximum clump masses used, 
are shown in Fig.~\ref{conti-susy}. 
The SUSY models that give a 5-$\sigma$ (or more) detection using
an Atmospheric Cherenkov Telescope (ACT)
with effective area $A_{eff}=10^{8} \rm{cm}^{2}$, energy threshold
$E_{th}=50$ GeV, and 100 hours of observation, are all the models that lie
above the corresponding line for each density profile. The backgrounds used to
calculate the noise are the hadronic and electronic cosmic ray shower contributions.  
Clearly, massive clumps appear to be easily detectable, 
regardless of profile; less massive clumps may be detectable, depending on their
degree of central concentration. For more details see \citet{tasitsiomi_olinto02}. 
There are some issues with respect to the  
ability of dark matter clumps to survive tidal disruption 
at distances small enough to yield
easily detectable fluxes.
In addition, there are some observational issues, given that the exact location of these clumps
is not known. Thus, we focus on an object whose location is known,
the LMC.  
\section{Flux from the LMC}
\label{sec:lmc}
We derive the density profile needed to calculate the $\gamma$-ray flux
using rotation velocity data \citep{kim_etal98,alves_nelson00}.
We fit the data using both the NFW ($\rho_{NFW}$), and the \citet{hayashi_etal03} profile ($\rho_{H}$), 
\begin{equation}
\rho_{NFW}=\frac{\rho_{0}}{r/r_{s} (1+r/r_{s})^{2}}\ ,\ \ \rho_{H}=\frac{\rho_{NFW}}{1+(r/r_{t})^{3}}\ \ . 
\end{equation}
Numerous observations indicate that the LMC is tidally stripped. 
The Hayashi et al. profile is a modification of the NFW which 
accounts for the  tidal stripping
that a halo may have  undergone.
\begin{figure}
\epsfig{file=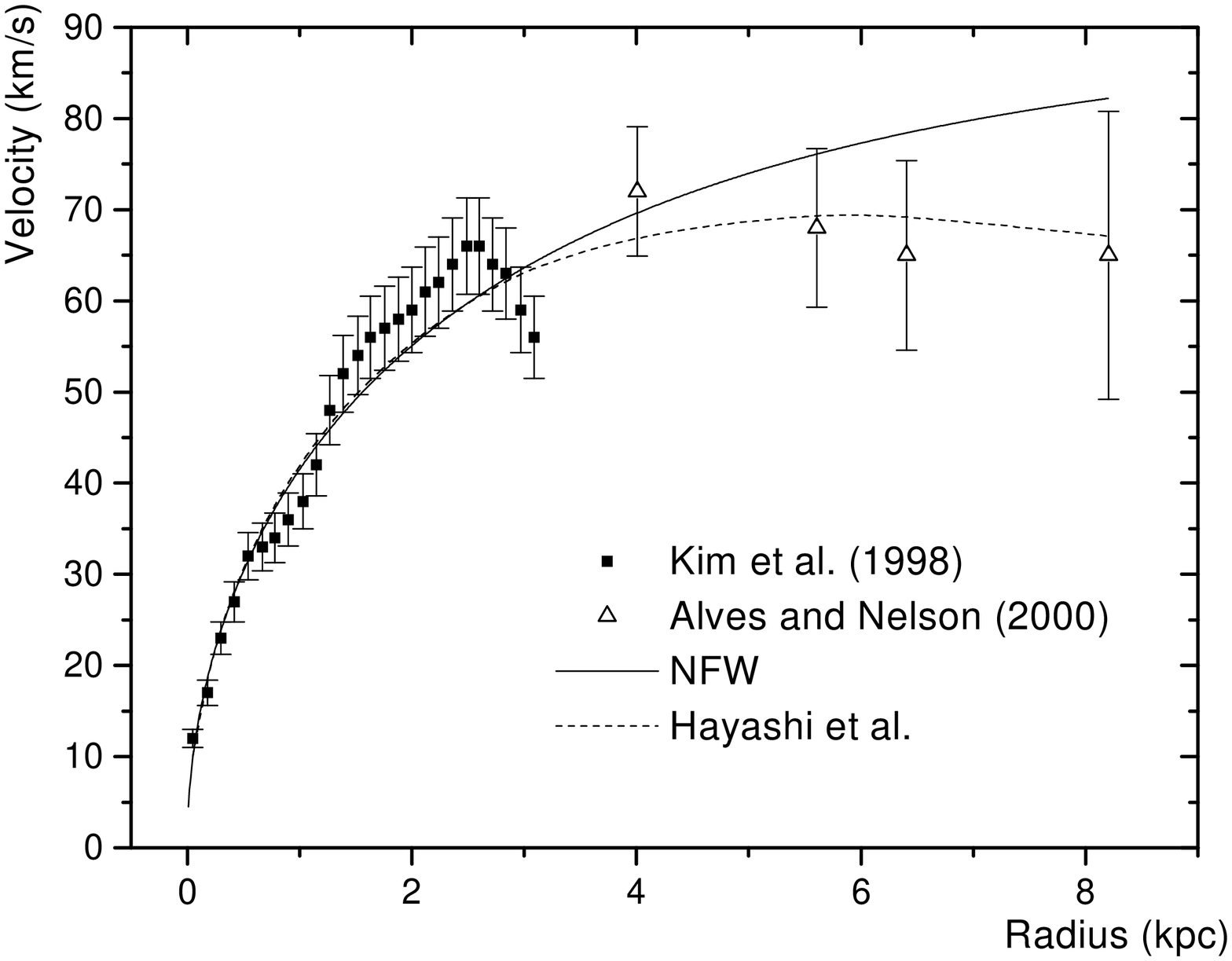, width=7.3cm, height=6.4cm}
\epsfig{file=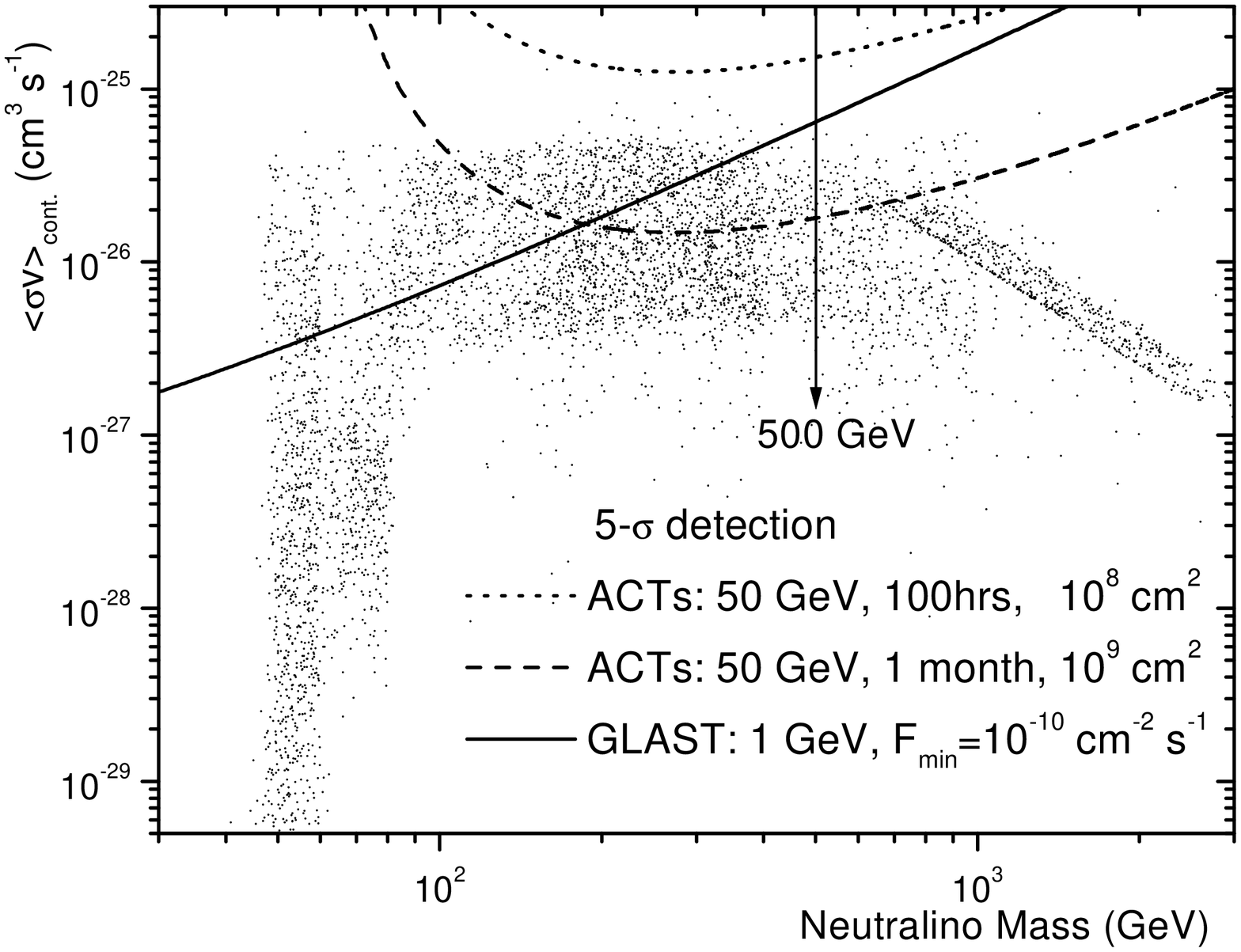, width=7.3cm, height=6.4cm}
\caption{\label{fig:lmc_susy}
Left panel: The LMC rotation curve (points) and the NFW and Hayashi et al. fits. 
Right panel: The minimum detectable $\langle \sigma v \rangle_{cont.}$  versus
$m_{\chi}$ for the Hayashi et al. profile (the NFW yields similar results).
Only  SUSY models that lie above the corresponding curve 
yield a detectable signal. The dotted line represents an 
observation feasible with upcoming ACTs. The 
dashed line assumes an effective area of $10^9$cm$^{2}$ which will be
achieved only at high energy thresholds ($\sim 1$ TeV); this along with the
relatively large integration time, renter this observation rather difficult 
to be achieved. The $F_{min}$ used for GLAST corresponds to one year of on-target 
observation and is the GLAST flux sensitivity for energies $\geq 1$ GeV. 
The vertical arrow at $m_{\chi}=500$ GeV corresponds to a recently derived upper limit on the
neutralino mass\citep{ellis_etal03}.}
\end{figure}
  
The fits are shown in the left panel of Fig.~\ref{fig:lmc_susy}.
In the right panel we present our results for the part
of the SUSY parameter space that gives a 5-$\sigma$ detection for GLAST and
a typical ACT; the instrument and observation parameters used to derive these
results are also shown. 
Requiring that $F_{LMC} \geq F_{min}^{GLAST}=10^{-10}$ cm$^{-2}$ s$^{-1}$ we find
that GLAST will be able to detect the signal for a significant part of the parameter
space. This is true especially if the recently derived limit $m_{\chi} < 500$
GeV is taken into account\citep{ellis_etal03}.
Assuming standard  specifications, 
ACTs will not be able to probe any part of the parameter space (dotted line).
Unless, 
large integration times (say, $\sim 1$ month) and effective areas 
(say, $\sim 10^{9}$ cm$^2$) are used (dashed line). Note though that
such integration times are fairly long for ACT observations, and that
such large effective areas for an energy threshold  $\sim$ 50 GeV are beyond
the goals of existing and upcoming ACTs.
These conclusions
remain essentially the same for all the profiles used to model the LMC halo.  
The spectrum and its characteristic features, such as
the cutoff at $E=m_{\chi}$ will be  useful in identifying  neutralino annihilation as 
the origin of the observed flux, 
especially in the case of GLAST where the prospects of detection are
fairly good. 
The monochromatic
lines produced by neutralino annihilation (e.g., the $\gamma \gamma$ line
at $E=m_{\chi}$) would be excellent observational signatures if the
cross sections for these  processes were 
not suppressed\citep{tasitsiomi_olinto02}. 

EGRET has detected a flux of  $(14.4 \pm 4.7) \times 10^{-8}$ 
photons $(E > 100\ \rm{MeV})$ cm$^{-2}$ $s^{-1}$ from 
the LMC \citep{hartman_etal99}. The 
emission due to neutralino annihilation can be anywhere from $\sim  
10^{-13}$ to $\sim 10^{-9}$  photons $(E > 100 \ \rm{MeV}) 
\ \rm{cm}^{-2} \ \rm{s}^{-1}$, depending on the neutralino parameters.  
The maximum possible flux is still $\sim$ 2 orders of magnitude  less than
the observed flux. This means that cosmic rays may be 
almost the exclusive source of the observed flux, as is often assumed. 
This is verified in the case of synchrotron emission from neutralino annihilation 
as well [for the synchrotron and for more details on the $\gamma$-rays 
see \citet{tasitsiomi_etal03}]. 
\section{Conclusions}
Dark matter clumps are in principle detectable, depending on the SUSY
parameters, their distance and their degree of central concentration. 
The expected $\gamma$-ray flux from the LMC  is, for most SUSY models, 
significantly  smaller 
than what EGRET observed, verifying the usual assumption
that  cosmic rays are almost exclusively the origin of the detected flux.
However, the flux is high enough to renter a large part of the SUSY parameter space accessible 
to GLAST; the detection of the signal by ACTs is highly unlikely.

\end{document}